# Rwandacoin: Prospects and challenges of developing a cryptocurrency for transactions in Rwanda

Oluwafunmilola Kesa, Violette Mahoro
Carnegie Mellon University
Kigali, Rwanda
{okesa,vmahoro}@africa.cmu.edu

*Abstract*— The use of cryptocurrencies such as Bitcoin and Ethereum in performing online transactions has been on the rise in the world. Africa as a continent is not left out in the adoption of blockchain and cryptocurrencies. This paper explores the prospects and challenges of developing a cryptocurrency in Rwanda which we denote Rwandacoin. In addition, the paper discusses the potentials of Rwandacoin easing intercountry trading in East Africa.

## I. Introduction

Many startups and initiatives are leveraging blockchain technology and cryptocurrencies in Africa. Blockchain is a public digital ledger in which transactions are recorded [6]. Transactions recorded on a blockchain are immutable i.e. cannot be modified. Blockchain uses a peer to peer network such that only one decentralized ledger is maintained. The immutability property of blockchain makes it a potential technology in various businesses and government services to ensure transparency.

Cryptocurrencies such as Bitcoin and Ethereum are built on the blockchain technology. A cryptocurrency is a virtual currency that uses cryptography to validate the owner of a unit of value of the currency [1]. Despite the rise of cryptocurrencies, governments are skeptical about accepting existing cryptocurrencies as legal method of payment. This skepticism is understandable as existing cryptocurrencies are volatile and have records of scams and illegal transactions. Thus, it is logical for a country to explore the intricacies of developing a cryptocurrency that is stable and regulated. This topic area is interesting because a stable and regulated cryptocurrency has many advantages including fewer paper notes in circulation, easier intercountry trading, and improved online transactions. In recent years, Rwanda has witnessed astonishing technological revolutions. Adopting a national cryptocurrency in Rwanda would improve its online transactions and evolve related technology innovations. In addition, various technology summits and forums including Next Einstein Forum (NEF), Transform Africa, and African Union Extraordinary Summit are held in Rwanda due to its friendly atmosphere. With the recent development of the African Continental Free Trade Area (AfCFTA) agreement [16], Rwanda could leverage a national cryptocurrency to promote easy trading with other African countries. If properly designed and harnessed, Rwandacoin could be widely accepted across the continent as the African cryptocurrency. This leads to an important question. What are the prospects and challenges of developing a regulated and stable cryptocurrency in Rwanda?

Our research focuses mainly on the prospects and challenges the government might face to develop a regulated and stable cryptocurrency for Rwanda. Our research does not involve developing the actual cryptocurrency. We assume that the government of Rwanda would be interested in cryptocurrency and there is no existing national cryptocurrency in Rwanda.

## II. Importance and prior work

Blockchain has a great potential in numerous fields including financial services, reputation systems, Internet of Things, and so on [2]. Despite this potential, there are many challenges to overcome in the usage of blockchain. Bitcoin, which was the first cryptocurrency built on blockchain, has challenges including scalability, privacy leakage, and selfish mining. Efforts to address scalability are storage optimization and redesigning the blockchain. Mixing and zerocoin are proposed to address privacy leakage. Lastly, selfish mining can be prevented by using random beacons and timestamps so that miners can select fresh blocks [2].

Although there are many blogs and news about countries that are considering or have developed their national cryptocurrencies [11, 13, 14, 17, 18, 19], only a few research articles have explored the prospects and challenges of developing one. The government of Rwanda might find this research useful in their consideration of the development of a regulated digital currency.

### A. Traditional Payment System versus Cryptocurrency Architecture

The conventional financial system always requires a third party i.e. central or commercial bank (see Fig. 1). The financial system allows government institutes to track money flow from one individual or company to another. An example is a case where Mr. X in Rwanda wants to transfer money from his Bank of Kigali account to Mr. Y's Maybank account in Malaysia. The transaction is reviewed by banks on both ends, unlike cryptocurrencies, these transactions disclose all information on both parties including names, gender, location and nationality. Fiat money is entirely controlled by the government or a regulatory body whose creation is modulated based on internal controls and policies, regulatory requirements or government backing the currency.

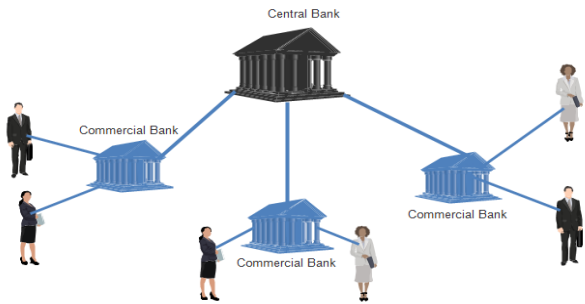

Fig. 1. Traditional System of Banking

The prominent feature in the design of the cryptocurrency architecture is the decentralized control which implies that no single authority, financial institution or legal office have control over the payment flow in the system. This architecture has received criticism for disrupting the conventional financial system where a third party, usually the central bank, oversees the movement of money from one individual or company to another. In Bitcoin's blockchain, for instance, blocks containing transactions are created through a proof-of-work mechanism [5]. The proof-of-work (PoW) mechanism ensures that nodes or miners on the network solve a puzzle before blocks can be created. This reduces conflicts in the system and limits the ability of attackers to control the system. The difficulty is set such that a new block is created in the network every 10 minutes on average [5]. This mechanism, although consumes both operational resources and capital, ensures the financial integrity and security of the decentralized system [4]. Other mechanisms, such as proof-of-stake, have emerged to make the blockchain technology less resource-intensive.

### B. Governments and Cryptocurrencies

Many governments currently ban Bitcoin operations in their countries, however, some of these countries are investigating the development of their own national cryptocurrencies. China which is home to about 71 percent of Bitcoin mining pools in the world recently intensified its efforts to limit Bitcoin mining [9, 10]. The government claims that Bitcoin mining consumes a lot of electricity in the country. Thus, the government is developing policies to curb the activities of miners. Despite the ban, the government of China is investigating the development of a national cryptocurrency. In Nigeria [8], the Securities and Exchange Commissions in 2017 warned Nigerians against investing in cryptocurrency as no regulations or guidelines have been developed to protect investors from financial losses. Similarly, the Central Bank of Nigeria advised banks to not transact virtual currencies in any way.

Recently, Venezuela launched the first state-backed cryptocurrency called Petro [12]. The Petro is based on blockchain technology and it's backed by the country's oil and mineral reserves. Ever since it was launched in February 2018, there has been different opinions about the credibility of the cryptocurrency. Some people believe that the Venezuelan governemt is using the cryptocurrency as a means to circumvent the US sanctions [13]. Although Venezuala has been in a severe economic crisis, the credibility or intentions of the cryptocurrency cannot be decided yet.

### C. Concerns with Cryptocurrencies

There are many concerns with cryptocurrencies including anonymity, stability, and regulation.

#### 1) Anonymity

Zerocoin is an extension to Bitcoin that provides strong anonymity. The authors [3] claim that Zerocoin is strong but it still reveals payments' destinations and amounts. The authors proposed a new currency called Zerocash which they believe is stronger than Zerocoin. Zerocash uses decentralized anonymous payment schemes (DAP schemes) which enables users to pay each other privately.

Despite user concerns, strong anonymity appears to be much of a challenge to law enforcers who face difficulty trying to minimize virtual money laundering. In [7], the authors analyzed anonymity of online transactions using Bitcoin. The authors' findings show that by analyzing transaction histories and using users' information supplied to Bitcoin exchanges, it may be possible to identify criminals. The probability of identifying criminals through this means is extremely low as most criminals do not give their real identity to Bitcoin exchanges.

#### 2) Stability

Due to high volatility, the market capitalization of the cryptocurrency industry is not stable. Many companies and banks reject integrating cryptocurrency into their payment systems because of fear. This fear is due to the sharp price changes of cryptocurrencies and difficulty in establishing a value. The right regulations could minimize the volatility of cryptocurrrency and make cryptocurrency appealing to the public.

#### 3) Regulation

Regulation is a major concern with cryptocurrencies. The emergence of cryptocurrencies has been a disruption for the government regulatory systems across the globe [14]. Governments have tight regulations and control over traditional banking systems which is not currently the case for the cryptocurrency framework. It has however been noted that the success of cryptocurrencies will depend on the way the regulatory framework works [14].

In addition, given the rise in interest and adoption of digital transactions and payment systems in almost every sector, and with cryptocurrencies gaining recognition worldwide, governments have, as well, expressed interest in developing policies and regulations that will allow this novel currency to expand. This would not only satisfy the privacy of the transactions but also allow government regulatory systems to mitigate the risks that come with the technology.

To develop Rwandacoin, a deep understanding of the Blockchain technology and the challenges experienced in existing cryptocurrencies is necessary. Here, we have reviewed some cryptocurrencies including Bitcoin, Zerocoin and Zerocash. The common concerns across these cryptocurrencies are stability, scalability, anonimity, and regulations.

### III. METHODOLOGY

#### A. Cryptocurrency Climate in Rwanda

With the common concerns in most countries for developing cryptocurrencies being scalability, anonymity and regulations, we gathered information regarding these concerns to understand the cryptocurrency climate in Rwanda. We conducted a survey at Carnegie Mellon University Africa and National Bank of Rwanda to analyze the feasibility and acceptance of the cryptocurrency.

#### B. Comparison with other currency-backed cryptocurrencies

To explore the prospects and challenges of developing Rwandacoin, we compared existing currency-backed cryptocurrencies. These cryptocurrencies include e-krona in Sweden, eDinar in Tunisia, and eCFA in Senegal.

### IV. RESULTS AND DISCUSSION

A survey was conducted in Rwanda with 45 responses. More than 70% of the respondents expressed concerns about cryptocurrencies and the major concern was stability (see Fig. 2). However, about 60% (TABLE I.) noted that a regulated cryptocurrency developed by the government would promote economic activities including online transactions. This shows that despite the concerns people have about cryptocurrencies, the likelihood of people using a national cryptocurrency is high (TABLE I.).

In addition, we looked at different countries that are interested in or have developed their national cryptocurrencies. In 2015, it was reported that Tunisia was replacing its digital currency with a blockchain-based version [21]. However, it is not clear whether the country developed the cryptocurrency or not. A cryptocurrency called eDinar exists on coinmarketcap [22] but there are no details to verify that it's in use in Tunisia. Further articles noted that the eDinar coin uses a delegated proof-of-stake (DPOS) which is claimed to have better data protection [23]. A delegated proof-of-stake can be thought of as a semi-centralized system where only a few elected stakeholders can create blocks [25]. The delegated proof-of-stake seem more suitable for the Rwandacoin architecture than the proof-of-work mechanism. However, further research is required to explore the safety of this architecture.

Another ongoing national cryptocurrency called eKrona has been under analysis and inquiry by Riksbank, the central bank of Sweden since 2017. The bank has identified no major obstacle for the coin and is currently investigating the possibility of leveraging digital infrastructures and collaborating with other actors to propose interactive solutions for end-users [20]. According to the project plan, the cryptocurrency would likely be implemented in 2019. The e-krona would be equivalent in value to the country's regular currency, kronor [11].

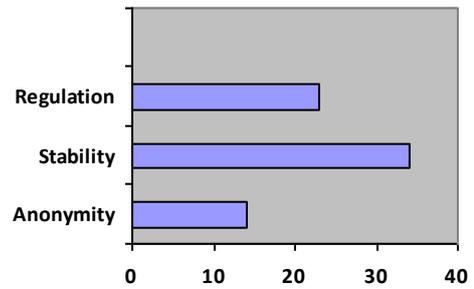

Fig. 2. Peoples' concerns with cryptocurrencies

TABLE I. SURVEY RESULTS

| Question | Yes | No | Maybe |
| --- | --- | --- | --- |
| Would you be willing to use government regulated cryptocurrency? | 44.4% | 20% | 35.6% |
| Would a regulated cryptocurrency promote economic activities (e.g online transactions, trading etc.) in Rwanda? | 60% | 13.3% | 26.7% |

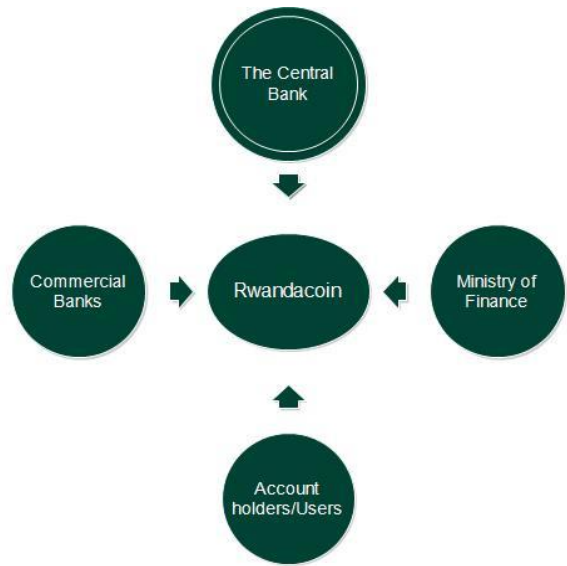

Fig. 3. Participants involved in the development of Rwandacoin

The eCFA is a digital currency based on blockchain, introduced by the Banque Regionale de Marches in Senegal. The Central Bank of West African States is the official regulator of the eCFA and has set rules in the member states on how the currency will be used [24]. However, the rules were not made known to the public and the state of the cryptocurrency is not known.

As suggested by the former US Federal Reserve Governor, Kevin Warsh, a central bank regulated cryptocurrency leveraging the blockchain technology would have significant influence on the national economic activities and as well address the current concerns of the non-regulated cryptocurrencies [15]. He also emphasized on the reduced money laundering as central banks build systems that push for law enforcement.

## V. CONCLUSION

Considering the ongoing development of cryptocurrencies in general, we see a lot of prospects for the Rwandacoin. However, there are challenges the government of Rwanda must take into consideration including the acceptability of the cryptocurrency by the public. Different participants in the country (see Fig. 3) should be involved to elicit the best approach for the design and development of the Rwandacoin. Further research should include the design of an architecture for a regulated Rwandacoin.